\documentclass[11pt,twoside]{article}
\usepackage{amsmath} 

\usepackage[english]{babel}
\usepackage{amsbsy}
\usepackage{amsfonts}
\usepackage{amssymb}
\usepackage[pdftex]{graphicx}
\usepackage{color}
\usepackage{float}
\usepackage{framed}
\usepackage{url}
\usepackage{amsmath}
\usepackage{color, colortbl}
\usepackage[table]{xcolor}
\usepackage{mathrsfs}

\usepackage[T1]{fontenc}
\usepackage[utf8]{inputenc}
\usepackage{authblk}

\newcommand\blfootnote[1]{%
	\begingroup
	\renewcommand\thefootnote{}\footnote{#1}%
	\addtocounter{footnote}{-1}%
	\endgroup
}

\title{\textbf{An Exact Solution for Allocating Car Parking Spaces on Campus}}
\author[1]{Luke O. Joel}
\author[2]{Sawyerr A. Babatunde}
\author[1]{Adewumi O. Aderemi}
\affil[1]{School of Mathematics, Statistics and Computer Science\\
           University of Kwazulu-Natal\\
           Westville Campus, Durban\\
           South Africa}
\affil[2]{Department of Computer Sciences\\
           University of Lagos\\
           Yaba\\
           Lagos, Nigeria}\date{}

\begin{document}


%

\maketitle

\pagestyle{myheadings}
\markboth{Luke O.J., Sawyerr B.A and Adewumi A.O.}{iSTEAMS Research, Nexus 2013}
\thispagestyle{empty}

\begin{abstract}

All over the world, especially in the university environment, planning managers and traffic engineers are constantly faced with the problem of inadequate allocation of car parking spaces to demanded users. Users could either prefer reserved parking spaces to unreserved parking spaces or vice versa. This makes the campus parking manager to be faced with two basic problem which are: the problem of allocating the actual number of available reserved spaces to users without any conflict over the same parking space, and the problem of determining the number of parking permit to be issued for parking lot with unreserved spaces. Hence, an optimal or available solution to the problem is required. This paper investigates a model for allocating car parking spaces, adds a constraint to address the reserved parking policy in a university environment and solves the parking allocation problem using an exact solution method. The result obtained gives the value of the objective function and the optimal allocation of users to each parking lot. \blfootnote{\textbf{An International Multidiscinary Conference on Research, Development and Practices in Science, Technology, Education, Arts, Management \& the Social Science (iSTEAMS) \\ Conference Centre, University of Ibandan, Nigeria. 30 May - 01 June 2013.}}
\end{abstract}

\textbf{Keywords:} Allocation, Model, Parking space, Parking lot, Reserved spaces, University

\section{Introduction}
Parking is a major concern in the transportation planning and traffic management of any organisation all over the world. Parking problems, among other things, are major problems facing the society and especially the university environment due to limited number of available parking spaces and the cost of parking facilities. The challenge is to develop a model of the problem that considers different parking policies in the campus environment and to obtain an optimal or available solution to the problem. Many studies have looked at the parking problem from administrative
and management point of view. However this paper will examine the problem from optimization point of view. The paper addresses the problem of parking allocation in the university environment by formulating a model of the problem which caters for both reserved and unreserved policy in the campus world

\section{Related Works}

In the previous studies on parking problem, Narragon, Dessouky and DeVor \cite{ref4} evaluated campus parking over-issuance policies by developing a probabilistic model which permits different classes of users to be considered simultaneously. Mouskos et al. \cite{ref3} formulated a deterministic dynamic parking reservation system (PRS) for performing parking space assignment on the minimization of parking cost in order to aid users in securing a parking space either before or during their trip. Chiu \cite{ref5} developed a multi-objective linear integer programming model for the optimum allocation of the off-street parking facilities decision makers. He advocated for the use of existing public facility as a parking facility. Batabyal et al. \cite{ref7} analysed two university parking issues by determining the mean parking time of an arriving car for both short term and long term parkers and computing their probability distribution function. He also calculated the probability distribution function of parking violators. Sattayhatewa et al. \cite{ref8} modelled the evaluation of parking lot choice by considering three (3) major factors- driving time, parking cost, and walking time which could be used to analyse the current traffic conditions, improve the traffic conditions and assess various operational and management policies for special events. Brown-West \cite{ref9} presented an optimization methodology for the use of existing land and to manage parking spaces in a competitive, policy-driven university campus. Major operational and site features, as well as parameters that could help parking managers and engineers are included in the model.
Essentially, Goyal and Gomes \cite{ref1} proposed a parking allocation model in a university environment on cases where the number of users is equal or less than the available spaces and where the number of users is greater than the available parking spaces.  The latter case will be the focus of this paper with reserved policy.

\section{Parking Allocation Model}
The allocation of available parking spaces to a set of users in order to minimize the distance walked by each user from the parking lot to the buildings in which they work is a difficult one, especially when the reserved policy is to be considered. A parking reserved policy is an important part of campus parking, hence there is need to incorporate this into the campus parking space allocation model. A constraint that addresses the reserved policy, Equation \ref{eq:pq8}, was introduced to the model proposed in \cite{ref1} for a case where the number of users, $T_U$, is greater than the available spaces, $T_S$. The constraint ensure that sum of the parking allocation, $X_{ijk}$, is equal to the number of available spaces, $M_{ik}$, for reserved policy and greater than the number of available spaces for unreserved policy.

  \begin{equation}
  \label{eq:pq8}
  \sum_{j=1}^{m} X_{ijk} \geq M_{ik} \hspace{5mm} \mbox{for} \hspace{2mm} i = 1,2, ... ,l \hspace{2mm} and \hspace{2mm} k = 1, 2, ..., n
  \end{equation}

By reserved spaces, we mean, a user given a reserved allocation does not share his allocation with any other user. That is, the number of users allocated to a parking space marked reserved cannot be more than the number of parking spaces. For unreserved, the number of users could be more than the available parking spaces because they are meant to be shared by more than one user. Hence, the model is formulated as an linear programming model and it is given as:

  \begin{equation}
  \label{eq:pq5}
  Minimize \hspace{5mm} Z = \sum_{k=1}^{n} \sum_{j=1}^{m} \sum_{i=1}^{l}D_{jk}X_{ijk} 
  \end{equation}
  
subject to:

  \begin{equation}
  \label{eq:pq6} 
  \sum_{k=1}^{n} X_{ijk} = P_{ij} \hspace{5mm} \mbox{for} \hspace{2mm} i = 1,2, ... ,l \hspace{2mm}and \hspace{2mm} j = 1, 2, ..., m
  \end{equation}

 \begin{equation}
  \label{eq:pq7}
  \sum_{j=1}^{m} \sum_{i=1}^{l} X_{ijk} = A_{k} \hspace{5mm} \mbox{for} \hspace{2mm}  k = 1,2, ... ,n
  \end{equation}

   \begin{equation}
    \label{eq:pq88}
    \sum_{j=1}^{m} X_{ijk} \geq M_{ik} \hspace{5mm} \mbox{for} \hspace{2mm} i = 1,2, ... ,l \hspace{2mm} and \hspace{2mm} k = 1, 2, ..., n
    \end{equation}

  \begin{equation}
  \label{eq:pq9}
  X_{ijk} \geq 0  \hspace{1cm} \forall \hspace{5mm} i,j,k \geq 0
  \end{equation}

\noindent Where, \\
$ l $ = the total number of permits type (with index $i$) \\
 $ m $ = the total number of users' building (with index $j$ ) \\
 $ n $ = the total number of parking lot (with index $k$) \\
 
  \begin{equation*} 
   T_S = \sum_{k=1}^{n} N_{k} = \sum_{k=1}^{n} \sum_{i=1}^{l} M_{ik} 
   \end{equation*}
 
 \noindent  $ T_S $ = the total number of available parking spaces \\
   $ N_{k} $ = the number of available spaces in the  $kth$ parking lot excluding the spaces for handicapped users \\
 $M_{ik}$ = the number of parking places available with permit type $ i$ in $kth$ parking lot \\
  $ A_{k} $ = the number of permits issued to the  $kth$ parking lot \\
   $ D_{jk} $ = the distance between the $jth$ users' building and the $kth$ parking lot \\
  $ X_{ijk} $ = the number of people having permit type $i$, users' building $j$ in the $kth$ parking lot \\
 
 \begin{equation*} 
        T_U = \sum_{i=1}^{l} B_{i} = \sum_{j=1}^{m} \sum_{i=1}^{l} P_{ij} 
 \end{equation*}
 
\noindent $ T_U $ = the total number of users demanding parking \\
 $ B_{i} $ = the number of permit type $i$ users  \\
  $ P_{ij} $ = the number of permit type $i$ users working in building $j$ \\

\noindent The objective function in Equation \ref{eq:pq5} minimizes the distances walked by users from each parking lot to their respective buildings. Equation \ref{eq:pq6} is the permit type users constraint which ensures that the sum of the parking allocation in each parking lot is equal to the number of users with the permit type for the parking lot. Since several parking permits are issued for different parking lot, the constraint in Equation \ref{eq:pq7} ensures that the sum of parking allocation for users with permit type $i$ working in building $j$ is equal to the parking permit issued for the parking lot. Equation \ref{eq:pq88} is the reserved spaces constraint introduced to the model and it is as explained earlier. The non-negativity constraint in Equation \ref{eq:pq9} keeps the variables to be equal or greater than zero.

\noindent \cite{ref1} made the following assumptions:
\begin{enumerate}
\item 	The shortest walking distance between each parking lot and the users' working building is known and it is taken by all users.
\item  The probability of a user bringing his car on a particular day and the probability of the user finding a space on that day is the same for all users.
\end{enumerate}

\section{Data}
 
The data used is from the parking data for University of KwaZulu-Natal (UKZN), Westville Campus \cite{ref6}.
 UKZN has a staff population of approximately 4300 people, and about 40\% of this were from Westville Campus, which is approximately 1720 people. Obviously, not all the 1720 people will need a parking space, so about 75\% of the Westville Campus staff require parking spaces, which gives us a population of 1290 users demanding parking spaces. \newline 
 A look at the available parking spaces is necessary for the efficient calculation of the ratio of demand to supply of parking spaces on Westville campus. Hence, from Table \ref{table:pt1}, we are to allocate 1047 parking spaces to 1290 users with some reserved consideration. There are several buildings and parking lots in Westville Campus but twelve(12) out of these buildings and six(6) out of these parking lots are used in the study. A break down of the users demanding parking in each of this building is given in Table \ref{table:pt2}. The distance cost from each building to each parking lot is calculated and given in Table \ref{table:pt3}.
 
 \begin{table}[htb]
 \caption{Available Parking Spaces in the Parking Lots}
 \centering
 \begin{tabular}{c c c c}
 \hline\hline
 Parking Lots & Number Available & Reserved & Unreserved\\ [0.5ex]
 \hline
 1 & 201 & 40 & 161 \\
 2 & 138 & 138 & - \\
 3 & 126 & 27 & 99 \\
 4 & 142 & 32 & 110 \\
 5 &  68 & 68   & -   \\
 6 & 372 & 72 & 300 \\ 
 \hline \hline
 Total & 1047 & 377 & 670 \\[1ex]
 \hline
 \end{tabular}
 \label{table:pt1}
 \end{table}

  \begin{table}[htb]
   \caption{Population of Users demanding parking}
   \centering
   \begin{tabular}{c c c c}
   \hline\hline
   Buildings & Users demanding parking &  Reserved Number & Unreserved Number\\ [0.5ex]
   \hline
   1 & 142 & 41 & 101 \\
   2 & 77 & 23 & 54\\
   3 & 118  & 34 & 84 \\
   4 & 64 & 19 & 45 \\
   5 &  60 &  19 & 41 \\
   6 & 220  & 64 & 156 \\ 
   7 & 51 & 15 & 36\\
   8 & 129 & 38 & 91 \\
   9 &  42 & 11 & 31\\
   10 &  103 & 30 & 73\\
   11  &  169 & 49 & 120\\
   12  &  115  & 34 & 81\\
   \hline \hline
   Total & 1290 & 377 & 913 \\[1ex]
   \hline
   \end{tabular}
   \label{table:pt2}
   \end{table}

   \begin{table}[htb]
      \caption{The Distance Cost}
      \centering
      \begin{tabular}{c c c c c c c}
      \hline\hline
      Building & \multicolumn{6}{c}{Parking Lots} \\
      \hline
         & 1 & 2 & 3 & 4 & 5 & 6 \\ [0.5ex]
      \hline
      1 & 255 & 270 & 440 & 165 & 285 & 610 \\
      2 & 150 & 165 & 335  & 60 & 180 & 505 \\
      3 & 165 & 180  & 320  & 75 & 195 & 490 \\
      4 & 120 & 135  & 275 & 120 & 150 & 445 \\
      5 &  270 & 285  &  260  & 105 & 200 & 430 \\
      6 & 180 & 150  & 215  & 195 & 60 & 485 \\ 
      7 & 60 & 90 & 320 & 150 & 180 &  490 \\
      8 & 90 & 60 & 290  & 165 & 150 & 400 \\
      9 &  350 & 320 & 210  & 260 & 245 & 90 \\
      10 &  440 &  410 & 120  & 350 & 335 & 180 \\
      11  &  320 &   290 & 60  & 230 & 215 & 200 \\
      12  &  335  &  305  & 75  & 245 & 230 & 215 \\[1ex]
      \hline
      \end{tabular}
      \label{table:pt3}
      \end{table}

  The mean and the standard deviation \cite{ref1} for the distribution is given as $ p.A_{k}$ and $ \sqrt{(p(1-p)A_{k})}$ ,  where $ p $  is the probability of a user bringing his car on a particular day. Goyal \& Gomes \cite{ref1} observed that in order to get equal probability for all the users, the equation in (\ref{eq:pq10}) must be satisfied  
  \begin{equation}
  \label{eq:pq10} 
  \frac{N_{k} - p.A_{k}}{\sqrt{(p(1-p)A_{k})}} = \Psi 
  \end{equation} 
  
 \noindent Getting the value of $ \Psi$ enables us to calculate the total number of permit issued for the $kth$ parking lot. However, In order to obtain the value of $\Psi$ that will be used to calculate the total number of permit issued, $A_{k}$, which will be equal to the number of users demanding parking, $T_U$, Goyal \cite{ref2} suggested squaring Equation (\ref{eq:pq10}), rearranging the terms for the $kth$ parking lot, ignoring small terms and then equating it to the total number of users demanding parking. The resultant equation is the quadratic equation in (\ref{eq:pq11}).
  \begin{equation}
  \label{eq:pq11}
  n\Psi^2 - 2\Psi \sum_{k=1}^{n} \sqrt{N_{k}} - 2(p.T_U - \sum_{k=1}^n N_{k}) = 0
  \end{equation}

\noindent Solving the quadratic equation in (\ref{eq:pq11}), the value of $\Psi$ obtained is used to get the values of $A_{k}$ in equation (\ref{eq:pq12}):
  
  \begin{equation}
  \label{eq:pq12}
     A_{k} = \frac{(2N_{k} + \Psi^2 ) - 2\Psi \sqrt{(N_{k})}}{2p}
  \end{equation}
  
  \subsection*{Parking Permit Calculation}
The numbers of parking permit to be issued for each parking lot are calculated. These numbers are only calculated for the parking lots that are not entirely for reserved parking, since the numbers of parking permit issued cannot be greater than the numbers of available parking spaces in a reserved parking.
Hence, the number of parking permit issued will not be  calculated for parking lot 2 and 5. We subtract the number of parking spaces in the two parking lots for reserved parking lots only - 206, from the total number of users demanding parking - 1290, to obtain the number of users - 1084 to be used in calculating the parking permit issued.
The values of the variables to be substituted into Equation(\ref{eq:pq11}) are  
\begin{align*}
B = 1290, N_{1} = 201, N_{2} = 138, N_{3} = 126, \\
 N_{4} = 142, N_{5} = 68, N_{6} = 372, n =4,  
\end{align*}

\noindent with $p=0.7$ as suggested by \cite{ref2, ref4}. \\
which gives 

\begin{align*}
     4\Psi^2 - 2\Psi (\sqrt{N_{1}} + \sqrt{N_{3}} + \sqrt{N_{4}}+\sqrt{N_{6}} ) \\ -2(0.7*1084 - (N_{1}+N_{3}+N_{4}+N_{6}))=0 
\end{align*}

and finally the quadratic equation 
\begin{align*}
6\Psi^2 - 113.2122\Psi + 164.4 = 0
\end{align*}
Solving the quadratic equation, we obtain the practical value of $ \Psi = 1.535  $ \newline

\noindent Hence, we put the value of $\Psi$ got into Equation (\ref{eq:pq12}) to get the parking permit issued for each parking lot. Table \ref{table:pt4} gives the detail of that. Parking Lots 2 and 5 are only for reserved parking, while the other parking lots have a certain number for reserved and unreserved parking as given in Table \ref{table:pt1} .

\begin{table}[htb]
\caption{Parking Permit Issued for each Parking Lot}
\centering
\begin{tabular}{c c c}
\hline\hline
Parking Lots($k$) & Number Available($N_{k}$) & Parking Permit Issued($A_{k}$)\\ [0.5ex]
\hline
1 & 201 & 258 \\
$\ast$2 & 138 & 138 \\
3 & 126 & 157 \\
4 & 142 & 178 \\
$\ast$5 &  68 & 68 \\
6 & 372 & 491 \\ 
\hline \hline
Total & 1047 & 1290\\[1ex]
\hline
\end{tabular}
\label{table:pt4}
\end{table}

\section{Results and Discussion}
\begin{figure}
\begin{center}
\includegraphics[scale=2,width=0.75 \textwidth]{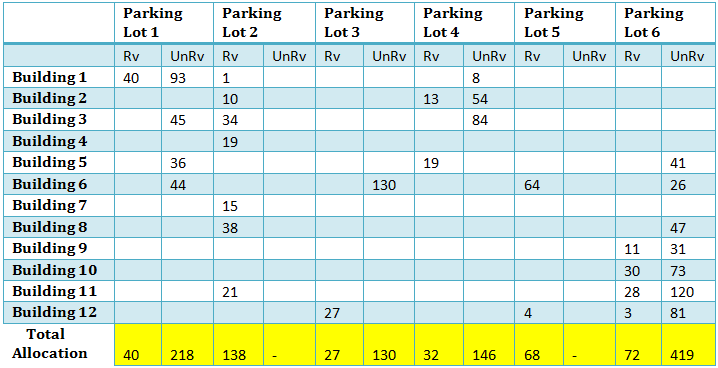}
\caption{Parking Allocation with Reserved Constraint}
\label{fig1}
\end{center}
\end{figure}

\begin{figure}
\begin{center}
\includegraphics[scale=2,width=0.75 \textwidth]{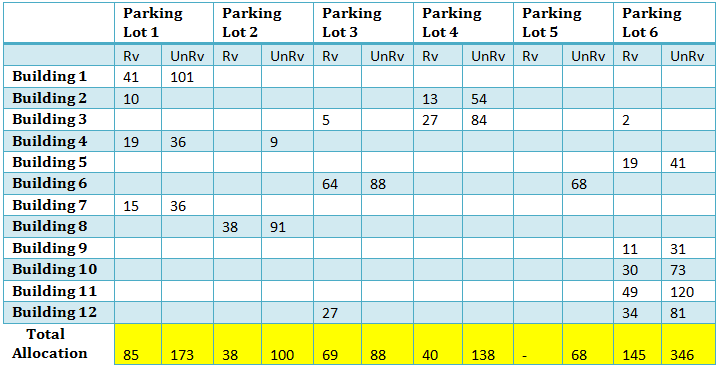}
\caption{Parking Allocation without Reserved Constraint}
\label{fig2}
\end{center}
\end{figure}

\begin{figure}
\begin{center}
\includegraphics[scale=2,width=0.6 \textwidth]{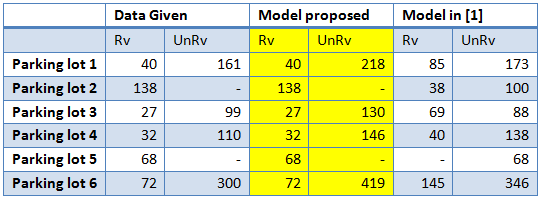}
\caption{Comparing the Allocation with the Data given}
\label{fig3}
\end{center}
\end{figure}

\begin{figure}
\begin{center}
\includegraphics[scale=2,width=0.6 \textwidth]{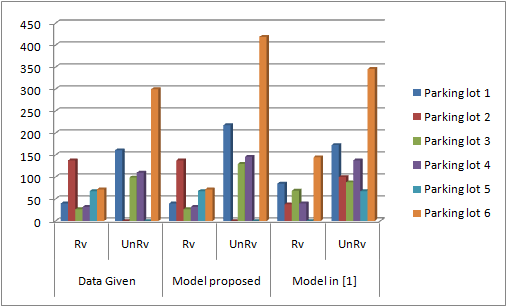}
\caption{Comparing the Allocation with the Data given 2}
\label{fig4}
\end{center}
\end{figure}

The mathematical model was implemented using IBM ILOG CPLEX software. The IBM ILOG CPLEX optimization studio (simply called CPLEX version 12.4) uses a variants of simplex method or the barrier interior point method to solve different kind of optimization problems. The CPLEX software package is incorporated into the Advanced Interactive Multidimensional Modeling System (AIMMS) which is used to obtain optimal solution to the problem. The following is the discussion of the results obtained by using the CPLEX software package. \\ \\
Figure \ref{fig1} gives the parking allocation for the formulated model with reserved constraint while Figure \ref{fig2} gives the parking allocation for the model formulated without reserved constraint in \cite{ref1}. Reserved space is abbreviated to 'Rv' and Unreserved space to 'UnRv' in the results shown in Figure \ref{fig1}, \ref{fig2}, \ref{fig3}, and \ref{fig4}. The value of the objective function, $Z$, in Equation (\ref{eq:pq5}), is the minimized value of the distances walked by the users from each parking lot to their respective buildings provided the constraints are satisfied. The objective value obtained for the allocation in Figure \ref{fig1} is $229160$ while that of the allocation in Figure \ref{fig2} is $210395$. Although the objective value of the allocation in Figure \ref{fig2} gives a minima value of both but the allocation obtained is infeasible based on the data. Comparing the two allocations, Figure \ref{fig1} indicates that the reserved parking spaces were allocated to reserved users. But in Figure \ref{fig2}, the allocation is contrary. That is, the number of users that are assigned to reserved parking in Figure \ref{fig2} are more than the number of reserved spaces available. Parking lot 2 and 5 are for reserved users, see Table \ref{table:pt1} and \ref{table:pt4}. This was emphasized by the parking allocation shown in Figure \ref{fig1} but not in Figure \ref{fig2}. Also, the number of spaces for reserved parking in parking lot 1,3,4, and 6 are the same with the number of allocated reserved users for these parking lots only in Figure \ref{fig1}.

Figure \ref{fig3} gives a close comparison of the total number of allocated spaces in each parking lot for the two models. Figure \ref{fig4} is a graph representing the same values as shown in Figure \ref{fig3}. The number of reserved and unreserved spaces allocated using the model with reserved constraints is highlighted in yellow. the results indicate that adding a constraint to address the reserved policy in a campus environment to the model formulated in \cite{ref1} is necessary in order to obtain feasible solution to the campus parking space allocation problem. In general, the allocation that was done assigned as much as possible the closest parking lot to users in a particular building while considering the interest of the remaining users in other building. Where the closest parking lot would not be possible, a little farther parking lot would be considered.

\section{Conclusion}

The model for allocating car parking spaces in the university with reserved constraint policy was investigated. An added constraint was introduced to the model proposed in \cite{ref1} so as to accommodate the reserved policy which is an important part of any university transportation planning.  Some parking data were use to test the model. An exact solution, the optimum objective function value and the allocation of users to each parking lot were obtained using the CPLEX software.

%

\end{document}